\documentstyle[12pt]{article}
\begin{document}
\begin{center}
{\bf Quantum superintegrable systems for arbitrary spin} \\

{\bf G.P.Pronko}\\

{\it Institute for High Energy Physics , Protvino, Moscow
reg.,Russia}\\
{\it Institute of Nuclear Physics, National Research Center
"Demokritos", Athens, Greece}

\end{center}

\begin{abstract}
 In \cite{PS} was considered the superintegrable  system which describes the
 magnetic dipole with spin $\frac{1}{2}$ (neutron) in the  field of linear current.
 Here we present its generalization for any spin which preserves superintegrability .
 The dynamical symmetry stays the same as
 it is for spin $\frac{1}{2}$.
\end{abstract}

\section{Introduction}

There exist few quantum systems where the degeneration of spectrum is
bigger when it follows from geometrical symmetry of the problem. The famous
examples of such systems are isotropic oscillator, Kepler problem, rotator
and some other which have no physical interpretation. This supplementary
degeneration of the spectrum arises due to dynamical symmetry (which
includes trivial geometrical). In this way the geometrical symmetry $SO(3)$
extends to the group $SU(3)$  in the case of of isotropic oscillator and to
the group $SO(4)$ in case of bound spectrum od Kepler problem. This kind of
systems also called maximally super-integrable. Their characteristic
property is that all finite trajectories are closed.   30 years ago we with
Stroganov had found another example of the physical system which possesses
supplementary degeneration of its spectrum due to existence of hidden
symmetry. The system describe the magnetic dipole with spin $\frac{1}{2}$
(neutron) in the field of line current. The obvious, geometrical symmetry
is the symmetry $SO(2)$ with respect to rotation around $z$-axis, the
direction of current (the translation along z is trivially separated).
Dynamical group in this case is $SO(3)$. Here we are speaking about the
symmetry for the negative part of the spectrum. For scattering states this
group changes as in the case of Kepler problem and becomes the other real
form of complex $SO(3)$, namely $SO(2,1)$ (or $E(2)$ for $E=0$).

The peculiarity of the system which we discovered is that it describes the
particle with spin, what was not know before.   The question which was
raised soon after is whether it possible to preserve dynamical symmetry for
the particles with higher spins. The answer up to now was negative in spite
of many attempts \cite{VGB}\cite{B-SBGV}\cite{V}. The failure of previous
considerations was because people wanted to preserve the interaction of the
spin particle with the external field which corresponded to intuitive
picture. But the truth is that particle with higher spin may interact not
only by its dipole magnetic moment. For example, the particle with spin $1$
acquires the possibility to have apart from dipole also quadruple
interaction, for spin $\frac{3}{2}$-- octuple interaction et cetera
\footnote {As a matter of fact, the importance of other interaction for
higher spins manifests itself also in case of Heisenberg magnetic, which is
integrable only if the interaction between spins is modified.}. Certainly,
this modification of interaction does not describe any longer an elementary
particle like neutron in magnetic field of the linear wire. In the same
time the emerged system could be useful for the description of trapped
ultra cold atoms in this field --- the subject was intensely discussed in
the literature \cite{SS}\cite{S}. Apparently atoms, being extended objects
will manifest its structure in inhomogeneous magnetic field through
interaction much more complicated when dipole interaction of neutron. As we
shall see below, the requirement of maximal super-integrability fixes the
form of interaction up to finite number of the parameters --- for spin $s$
the number of parameters is $2s+1$. Accidentally or not, but almost all
known maximally super-integrable systems have a wide physical application.
The system which we considered in \cite{PS} was discussed in \cite{I} in
connection with the trapping of ultra-cold neutrons.  So it could happened
that the interaction we found for higher spins also includes some important
cases.

The Hamiltonian of the system, considered in \cite{PS} is given by
\begin{equation}\label{1}
{\cal H}=\frac{ p_{x}^2+p_{y}^2}{2m}-{\bf \mu H},
\end{equation}
where $\bf \mu$ is magnetic moment of the particle and $\bf H$ is
magnetic field of linear current directed along the $z$-axis:
\begin{equation}
{\bf H}=CI(\frac{y}{r^2},-\frac{x}{r^2}).
\end{equation}
The constant coefficient $C$ depends on the unit system, in
practical system $C=0.2$. Thus the final form of the Hamiltonian
will be
\begin{equation}\label{3}
{\cal H}=\frac{ p_{x}^2+p_{y}^2}{2m}-k\frac{s_{x}y-s_{y}x}{r^2},
\end{equation}
where the coefficient $k$ collected all constants. For spin
$\frac{1}{2}$ the spin operator proportional to Pauli matrices. The
Hamiltonian (\ref{1}) is invariant with respect to rotations around
$z$-axis generated by $J_z=L_z+s_z$. In addition to this geometrical
integral the Hamiltonian (\ref{3}) possesses two non-trivial :
\begin{eqnarray}\label{4}
A_x=\frac{1}{2}(J_3 p_x+p_x
J_3)+km\frac{s_{x}y-s_{y}x}{r^2}y\nonumber\\
A_y=\frac{1}{2}(J_3 p_y+p_y J_3)-km\frac{s_{x}y-s_{y}x}{r^2}x
\end{eqnarray}
The integrals (\ref{4}) together with Hamiltonian and $J_z$ form the
following algebra:
\begin{eqnarray}\label{5}
&[J_z,A_x]=iA_y, \quad [J_z,A_y]=-iA_x, \quad [A_x,A_y]=-i{\cal
H}J_z \nonumber\\
&[A_x,{\cal H}]=0, \quad [A_y,{\cal H}]=0.
\end{eqnarray}
If we define now the operators
\begin{equation}\label{6}
J_x=A_x(-{\cal H})^{-1/2}, \qquad  J_y=A_y(-{\cal H})^{-1/2},
\end{equation}
then the following commutation relations of $SO3$ algebra hold true:
\begin{equation}\label{7}
[J_i,J_j]=i\epsilon_{ijk}J_k
\end{equation}
While we designed  the operators $J_i$ we  had in mind the the
discrete spectrum, for which energy is negative. For positive energy
the algebra will be $SO(2,1)$, because some signs in (\ref{7}) will
change. The Casimir operator of the algebra (\ref{7}) is expressed
via Hamiltonian:
\begin{equation}\label{8}
{\bf J^2}=J_1^{2}+J_2^{2}+J_3^{2}=-\frac{1}{4}-\frac{m k^2}{2{\cal
H}},
\end{equation}
therefore the Hamiltonian is given by
\begin{equation}\label{9}
{\cal H}=-\frac{mk^2}{2}\frac{1}{{\bf J^2}+\frac{1}{4}}.
\end{equation}
The representations of $SO(3)$ characterized by integer or
half-integer spin. In our problem it is clear that the eigenvalues
of $J_3$ could be only half-integer due to addition of spin
$\frac{1}{2}$ and integer orbital momentum, therefore only
half-integer representations will be realized. So the eigenvalues of
${\bf J^2}$ will be $\frac{2n+1}{2}(\frac{2n+1}{2}+1)$, $n=0,1...$
and the spectrum of energy  will be
\begin{equation}
E_n=-\frac{mk^2}{2}\frac{1}{(n+1)^2}
\end{equation}
The supplementary degeneration in this case means that the spectrum does
not depends on the eigenvalue of $J_z$.

The existence of additional integrals of motion in this case based
completely on the properties of Pauli matrices which represent spin
$\frac{1}{2}$ operators for and direct substitution instead of it,
the matrices which represent any other spin immediately destroys the
whole construction. The exit of this situation we will discuss in
the next section.

\section{High spins}

Let us consider the quantum system which describes neutral particle with
arbitrary spin $s$ in the field of rectilinear electric current. The
general form of the  Hamiltonian in this case is given by the following
equation (here we omitted the trivial part of kinetic motion along the line
of the current):
\begin{equation}\label{11}
{\cal H}=\frac{{\bf p^2}}{2m}+\frac{M({\bf s,x})}{{\bf x^2}}
\end{equation}
where ${\bf p,x}$ are 2-dimensional vectors, ${\bf s}$ is spin operator
${\bf s}=(s_x,s_y,s_z)$. The matrix $M({\bf s,x})$ will be specified later.
Now we shall impose on $M({\bf s,x})$ only one condition
\begin{equation}\label{12}
[M({\bf s,x}),J_z]=0,
\end{equation}
where $J_z=L_z+s_z$. Now let us look for the additional integrals of
motion in the following form:
\begin{equation}\label{13}
A_i=\frac{1}{2}(p_i J_z+J_z p_i)+V_i({\bf s,x}),
\end{equation}
where $V_i({\bf s,x})$ is an operator which should be defined by
interaction $M({\bf s,x})$. The commutator of the first term of (\ref{13})
with Hamiltonian (\ref{11}) gives
\begin{equation}\label{14}
[{\cal H},\frac{1}{2}(p_i J_z+J_z p_i)]=\frac{i}{2}\{J_z,\partial_i
\frac{M({\bf s,x})}{{\bf x^2}}\},
\end{equation}
where $\{A,B\}=AB+BA$. This form of commutator suggests the following
structure of the operator  $V_i({\bf s,x})$ :
\begin{equation}\label{15}
V_i({\bf s,x})=\epsilon_{ij}\frac{x_i}{{\bf x^2}}M({\bf s,x})
\end{equation}
where $\epsilon_{ij}$ -antisymmetric tensor. Indeed, commuting
(\ref{15}) with Hamiltonian we obtain:
\begin{eqnarray}\label{16}
[{\cal H},\epsilon_{ij}\frac{x_i}{{\bf x^2}}M({\bf
s,x})]=-\frac{i}{2m}[-\left\{L_z ,\partial_i \frac{M({\bf
s,x})}{{\bf x^2}} \right\}\nonumber\\
+\left\{p_k,\epsilon_{ik}(\frac{M({\bf s,x})}{{\bf
x^2}}+x_j\partial_j\frac{M({\bf s,x})}{{\bf x^2}})\right\}],
\end{eqnarray}
 Now if we add and subtract $s_z$ to $L_z$ we
can rewrite (\ref{16}) in the following form:
\begin{eqnarray}\label{17}
[{\cal H},\epsilon_{ij}\frac{x_i}{{\bf x^2}}M({\bf
s,x})]=-\frac{i}{2m}[-\{J_z,\partial_i \frac{M({\bf s,x})}{{\bf
x^2}}\}+\partial_i\{s_z,\frac{M({\bf s,x})}{{\bf x^2}}\} \nonumber\\
+\left\{p_k,\epsilon_{ik}(\frac{M({\bf s,x})}{{\bf
x^2}}+x_j\partial_j\frac{M({\bf s,x})}{{\bf x^2}})\right\}].
\end{eqnarray}
Imposing on matrix $M({\bf s,x})$ apart from (\ref{12}) the
conditions
\begin{eqnarray}\label{18}
s_z M({\bf s,x})+M({\bf s,x})s_z=0,\nonumber\\
(\frac{M({\bf s,x})}{{\bf x^2}}+x_j\partial_j\frac{M({\bf
s,x})}{{\bf x^2}})=0,
\end{eqnarray}
we arrive at the commutativity of
\begin{equation}\label{19}
A_i=\frac{1}{2}(p_i J_z+J_z p_i)-m \epsilon_{ij}\frac{x_i}{{\bf
x^2}}M({\bf s,x})
\end{equation}
with Hamiltonian. Note, that the matrix $M({\bf s,x})$, which we had
in the previous section for spin $\frac{1}{2}$ satisfies both
conditions (\ref{18}) and it was the reason why we achieved the
commutativity of integrals (\ref{4}) with Hamiltonian. Now it is
possible to prove that the commutation relations for the components
of $A_i$ are
\begin{equation}\label{20}
[A_i,A_j]=-i\epsilon_{ij}J_z 2m{\cal H},
\end{equation}
provided the same conditions (\ref{12}) and (\ref{18}) are satisfied.

Now let us take care of matrix $M({\bf s,x})$. The second equation
(\ref{18}) is rather simple, it requires $M({\bf s,x})$ to be a
homogenous function of $x_i$ of degree 1. So we can present $M({\bf
s,x})$ in the form
\begin{equation}\label{21}
M({\bf s,x})=|{\bf x}|\mu({\bf s,n}), \qquad {\bf n}=\frac{{\bf
x}}{|{\bf x}|},
\end{equation}
where the matrix $\mu({\bf s,n})$ commutes with $J_z$ and
anticommutes with $s_z$. Let us consider these conditions in the
basis $|s,k>$ of the unitary representation of spin $s$. This basis
is defined by
\begin{eqnarray}\label{22}
&s_z |s,k>=k|s,k>, \qquad {\bf s^2}|s,k>=s(s+1)|s,k>\nonumber\\
&s_{+}|s,k>=\sqrt{s(s+1)-k(k+1)}|s,k+1>,\nonumber\\
&s_{-}|s,k>=\sqrt{s(s+1)-k(k-1)}|s,k-1>\nonumber\\
&k=s,s-1,...,-s.
\end{eqnarray}
In this basis  $\mu({\bf s,n})$ has its matrix elements
$\mu_{kk'}({\bf n})$
\begin{equation}\label{23}
\mu_{kk'}({\bf n})=<s,k|\mu({\bf s,n})|s,k'>.
\end{equation}
The first equation (\ref{18}) implies the following:
\begin{equation}\label{24}
(k+k')\mu_{k k'}({\bf n})=0.
\end{equation}
The solution of this equation is
\begin{equation}\label{25}
\mu_{k k'}({\bf n})=\delta_{k,-k'}a_k({\bf n}),\qquad a_{k}^{*}({\bf
n})=a_{-k}({\bf n}),
\end{equation}
where the last condition guaranties that $\mu({\bf s,n})$ will be
hermitian. Now let us impose the condition (\ref{12}) on matrix
$\mu({\bf s,n})$ :
\begin{equation}\label{26}
[J_z,\mu({\bf s,n})]=0\Rightarrow [L_z,a_k({\bf n})]+2ka_k({\bf
n})=0
\end{equation}
This equation fixes the $\bf n$-dependence of $a_k({\bf n})$:
\begin{equation}\label{27}
a_k({\bf n})=\alpha_k e^{-2ik\varphi}, \qquad
e^{i\varphi}=n_1+in_2,\qquad\alpha_k^{*}=\alpha_{-k}.
\end{equation}
So, the final expression for matrix $\mu_{k k'}({\bf n})$
\begin{equation}\label{28}
\mu_{k k'}({\bf n})=\delta_{k,-k'}\alpha_k e^{-2ik\varphi}
\end{equation}
contains $2s+1$ real parameters which define the set of $\alpha_k$. The
matrix $\mu_{k k'}({\bf n})$ could also be expressed in terms of operators
$\bf s$:
\begin{eqnarray}\label{29}
\mu({\bf s,n})=\left(\beta_s
(s_{+}n_{-})^{2s}+h.c.\right)+\left(\beta_{s-1}(s_z-s)(s_{+}n_{-})^{2s-2}(s_z+s)+h.c\right)\nonumber\\
+\left(\beta_{s-2}(s_z-s)(s_z-s+1)(s_{+}n_{-})^{2s-4}(s_z+s)(s_z+s-1)+h.c\right)...
\end{eqnarray}
The structure of this expression could be understood from the following
explanations. First, note that due to condition (\ref{26}), matrix
$\mu({\bf s,n})$ can depend only of the combinations of
$(s_{+}n_{-}),(s_{-}n_{+})$ and $s_z$. Second, matrix $\mu({\bf s,n})$ in
the representation (\ref{22}) is anti-diagonal and in order to obtain
operator which has non zero matrix elements in the upper and lower corners
we need to take a liner combination of $(s_{+}n_{-})$ and its conjugated in
maximal power--- for spin $s$ it is $2s$. In this way we obtain the first
term of (\ref{29}). The second term is obtained with the same strategy but
here we need to eliminate the action of $(s_{+}n_{-})^{2s-2}$ on the
vectors $|-s,s>$ and $<s,s|$. This is the reason of appearing the fringe
multipliers $(s_z-s)$ and $(s_z+s)$. The rest is just repetition of this
procedure. The parameters $\beta_k$ in (\ref{29}) play the same role, as
$\alpha_k$ in (\ref{27}) but only $\beta_s=\alpha_s$, the other are
different because of additional multipliers, depending on $s_z$ in
(\ref{29}).

It is interesting that  even for $s=\frac{1}{2}$, we have not only one type
of interaction, which respects dynamical symmetry, but two. Indeed,
according to present consideration the Hamiltonian
\begin{equation}
{\cal H}=\frac{{\bf
p^2}}{2m}+a\frac{s_{x}y-s_{y}x}{r^2}+b\frac{s_{x}x+s_{y}y}{r^2}
\end{equation}
also possesses dynamical symmetry. The additional term in this Hamiltonian
describes electric dipole in electric field $\frac{\vec r}{r^2}$ which is
produced by rectilinear charge.

Last issue which we are going to discuss is the analogue of the formula
(\ref{9}) in the generic case. Defining as in (\ref{6}) the operators
$J_i$, having in mind discrete spectrum we obtain
\begin{equation}\label{31}
{\bf J^2}+\frac{1}{4}=-\frac{m}{2}\frac{\mu({\bf s,n})^2}{{\cal H}}.
\end{equation}
Matrix  $\mu({\bf s,n})$ commutes with Hamiltonian, as it follows from it
definition and the form of ${\cal H}$ (\ref{11}) and from (\ref{31}) we
obtain
\begin{equation}
{\cal H}=-\frac{m}{2}\frac{\mu({\bf s,n})^2}{{\bf J^2}+\frac{1}{4}}.
\end{equation}
As matrix $\mu({\bf s,n})$ in the bases $|s,k>$ is anti-diagonal, it square
is diagonal
\begin{equation}
(\mu({\bf s,n})^2)_{k
k'}=diag\{|\alpha_s|^2,|\alpha_{s-1}|^2,...,|\alpha_s|^2\},
\end{equation}
so it could be written as a linear combination of projectors on the states
with definite values of $s_z$.

In conclusion we summarize the above discussion. It is shown that the
problem, introduced in  \cite{PS} for spin $1/2$ admits generalization for
arbitrary spin which preserves dynamical symmetry. The interaction depends
on $2s+1$ parameters which leaves a big freedom for applications. The
energy spectrum has the same character $1/n^2$ as in the case of spin$1/2$
, but for the wave functions we need to select the proper representations
of $SO(3)$ corresponding to given value of spin. In \cite{PS} we have
constructed explicitly invariant form of Schrodinger equation for our
system as it was done by Fock \cite{Fock} for Kepler problem. This form
exists  also for arbitrary spin. In the present paper we did not touche the
subject of possible applications of the system we considered, as this is
the theme for separate paper.

\section*{Acknowledgments}
The author is grateful to professors A.K.Likhoded and A.V.Razumov
for their comments and fruitful discussions.   This work  was
supported by the programme ENTER-2004/04EP-48,
 E.U.-European Social Fund(75{\%}) and Greek Ministry of development-GSRT (25{\%})
and by RFFI grant 07-01-00234.


\begin{thebibliography}{20}


\bibitem{PS}
G.P.Pronko, Yu.G.Stroganov, Zh.E.T.F v.72, p. 2048, 1977, (Soviet Physics
JETP, v.45, p.1072, 1997)
\bibitem{VGB}
L.Vestergaard Hau, J.A.Golovchenko, M.M. Burns, Phys.Rev.Lett., v. 74,p.
3138,1995
\bibitem{B-SBGV}
K.Berg-Sorensen,  M.M. Burns, J.A.Golovchenko, L.Vestergaard Hau,
Phys.Rev.A,v.53, p.1653, 1996
\bibitem{V}
A.I.Voronin, Phys.Rev.A, v.43, p.29, 1991
\bibitem{SS}
J.Schmiedmayer, A.Scrinzi, Quantum Optics, v.8, p.693, 1996
\bibitem{S}
J.Schmiedmayer, Phys.Rev.A, v.52, R13, 1995
\bibitem{I}
V.K.Ignatovich, Uspekhi Fizicheskikh Naul, v.166, p.303, 1996
\bibitem{Fock}
V.A.Fock, Zs.Phys., v.98, p. 145, 1935
\end{thebibliography}
\end{document}